\documentclass[iop, apjl]{emulateapj}
\usepackage{natbib}
\usepackage{graphicx}

\def\kms {{\mathrm{km}\,\mathrm{s}^{-1}}}

\defcitealias{Pereira:2012spic}{\mbox{Paper I}}

\def\CII{\ion{C}{2}}
\def\CaII{\ion{Ca}{2}}
\def\CaH{\CaII\, H}

\def\MgII{\ion{Mg}{2}}
\def\MgIIk{\ion{Mg}{2}\,k}

\def\SiIV{\ion{Si}{4}}
\def\hk{{h\&k}}
\def\HeII{\ion{He}{2}}
\def\MgIIhk{\MgII\, \hk}

\usepackage{mathptmx} 
\usepackage[colorlinks=true,urlcolor=blue,citecolor=blue,linkcolor=blue]{hyperref}

\shorttitle{An \emph{IRIS} View on Spicules.}
\shortauthors{Pereira et al.}

\begin{document}

\title{An \emph{Interface Region Imaging Spectrograph} first view on Solar Spicules}
  
   \author{T. M. D. Pereira$^{1}$}
   \author{B. De Pontieu$^{1, 2}$}%
   \author{M. Carlsson$^{1}$}%
   \author{V. Hansteen$^{1}$}
   \author{T. D. Tarbel $^{2}$}
   \author{J. Lemen$^{2}$}
   \author{A. Title$^{2}$}
   \author{P. Boerner$^{2}$}
   \author{N. Hurlburt$^{2}$}
   \author{J. P. W\"ulser$^{2}$}
   \author{J. Mart\'inez-Sykora$^{2,3}$}
   \author{L. Kleint$^{2,3,4}$}
   \author{L. Golub$^{5}$}
   \author{S. McKillop$^{5}$}
   \author{K. K. Reeves$^{5}$}
   \author{S. Saar$^{5}$}
   \author{P. Testa$^{5}$}
   \author{H. Tian$^{5}$}
   \author{S. Jaeggli$^{6}$}
   \author{C. Kankelborg$^{6}$}

\affil{$^1$ Institute of Theoretical Astrophysics, University of Oslo, P.O. Box 1029 Blindern, N--0315 Oslo, Norway}
\affil{$^2$ Lockheed Martin Solar and Astrophysics Laboratory, 3251 Hanover Street, Org. A021S, Bldg. 252, Palo Alto, CA 94304, USA}
\affil{$^3$ Bay Area Environmental Research Institute, 596 1st St West, Sonoma, CA 95476, USA}
\affil{$^4$ NASA Ames Research Center, Moffett Field, CA 94035, USA}
\affil{$^5$ Harvard-Smithsonian Center for Astrophysics, 60 Garden Street, Cambridge, MA 02138, USA}
\affil{$^6$ Department of Physics, Montana State University, Bozeman, P.O. Box 173840, Bozeman, MT 59717, USA}
\date{Received; accepted}

\begin{abstract}
Solar spicules have eluded modelers and observers for decades. Since the discovery of the more energetic type~II, spicules have become a heated topic but their contribution to the energy balance of the low solar atmosphere remains unknown. 
Here we give a first glimpse of what quiet Sun spicules look like when observed with NASA's recently launched \emph{Interface Region Imaging Spectrograph} (IRIS). 
Using IRIS spectra and filtergrams that sample the chromosphere and transition region we compare the properties and evolution of spicules as observed in a coordinated campaign with \emph{Hinode} and the \emph{Atmospheric Imaging Assembly}. %
Our IRIS observations allow us to follow the thermal evolution of type II spicules and finally confirm that the fading of \CaH\ spicules appears to be caused by rapid heating to higher temperatures. The IRIS spicules do not fade but continue evolving, reaching higher and %
falling back down after $500-800$~s.  
\CaH\ type II spicules are thus the initial stages of violent and hotter events that mostly remain invisible in \CaH\ filtergrams. These events have very different properties from type I spicules, which show lower velocities and no fading from chromospheric passbands.
The IRIS spectra of spicules show the same signature as their proposed disk counterparts, reinforcing earlier work. Spectroheliograms from spectral rasters also confirm that quiet Sun spicules originate in bushes from the magnetic network. Our results suggest that type II spicules are indeed the site of vigorous heating (to at least transition region temperatures) along extensive parts of the upward moving spicular plasma.
\end{abstract}


\section{Introduction}                          \label{sec:introduction}
A better understanding of solar spicules is essential to unlock the mysteries of the chromosphere and low solar atmosphere. These dynamic jet-like features dominate the solar limb and since their first reports they have been challenging to explain, as attested in various reviews \citep{Beckers:1968, Sterling:2000, Tsiropoula:2012}. Over the last decades spicules have been observed in a variety of filters and instruments \citep[e.g][]{Dere:1989, Wilhelm:2000, Zachariadis:2002, OShea:2005}, but the contribution of the \emph{Hinode} mission \citep{Kosugi:2007} has been nothing short of revolutionary. From high quality time series of spicules in the \CaH\ band of the \emph{Solar Optical Telescope} \citep[SOT,][]{Tsuneta:2008, Suematsu:2008} it emerged that there are at least two types of spicules \citep{DePontieu:2007}. Type I spicules have a slower rise and fall ($15-40\;\kms$) and long lifetimes ($3-10$ minutes), with a parabolic motion in space-time diagrams, while type II spicules have a faster upward rise ($30-110\;\kms$) followed by a fading from the \CaH\ passband after $50-150$~s; they are not observed to fall back down \citep[][hereafter \citetalias{Pereira:2012spic}]{DePontieu:2007, Pereira:2012spic}. Type II spicules are the most abundant (seen in quiet Sun and coronal holes), while type I spicules appear mostly in and around active regions.

The disappearance of type II spicules from the \CaH\ passband has fueled speculation that these may be heated to higher temperatures  \citep{DePontieu:2009, McIntosh:2009,  McIntosh:2010, Tian:2011}, possibly even to transition region (TR) or coronal temperatures as suggested by \citet{DePontieu:2011}. However, establishing the thermal evolution of spicules to the higher chromosphere and TR has been difficult because the spatial and temporal resolution of previous instruments is just barely enough to resolve their rapid motions \citep[see also][]{Pereira:2013spiclett}.
The recently launched IRIS mission \citep{IRIS-paper} is aimed precisely at observing the dynamic interface between the chromosphere and corona, therefore making it ideal for the study of spicules. The aim of this work is to provide an overview of the quiet Sun spicule dynamics and thermal evolution as observed with IRIS, comparing them with the \CaH\ spicules from \emph{Hinode}/SOT  and their higher temperature counterparts from the \emph{Atmospheric Imaging Assembly} \citep[AIA,][]{Lemen:2012}.

\section{Observations and Analysis}            \label{sec:obs}

We make use of a series of coordinated quiet Sun observations obtained with IRIS, \emph{Hinode}/SOT and AIA, taken on 2014 February 21. From IRIS we use the chromospheric and TR slit-jaw filtergrams at 279.6~nm (dominated by \MgII, at $\approx 10$~kK) and 140.0~nm (dominated by \SiIV, at $\approx 80$~kK under ionization equilibrium conditions), taken with a 19~s cadence. From SOT we use the 396.85~nm filtergrams (dominated by \CaH, at $\approx 9$~kK) from the Broadband Filter Imager, taken with a 4.8~s cadence. From AIA we use the 30.4~nm filtergrams (dominated by \HeII, at $\approx 100$~kK), taken with a 12~s cadence. The target of the observations was the quiet Sun at the south pole (no coronal hole was visible), and the time series duration was 91 minutes. The IRIS field of view was centered at $(x, y) = (6\farcs7, -965\farcs5)$.

We made use of IRIS calibrated level 2 data \citep[for details on the reduction see][]{IRIS-paper}. The SOT images were reduced using the same procedure as in \citetalias{Pereira:2012spic}. For AIA we used calibrated level 1.5 images \citep[see][]{Lemen:2012}. In all IRIS and SOT filtergrams shown we applied a radial density filter \citep[see][]{DePontieu:2007} to enhance the visibility of spicules. The IRIS, SOT, and AIA spatial resolutions are approximately $0\farcs33$, $0\farcs2$, and $1\farcs5$, respectively. The SOT and AIA images were aligned and interpolated to the IRIS pixel size of $0\farcs166$~pix$^{-1}$. 

Additionally, we make use of a very large dense raster observed with IRIS at the solar north pole on 2013 October 9 at 13:10 UT. This raster has 400 positions observed with a spatial spacing of $0\farcs35$ and a cadence of 9~s. The raster covered about $140\arcsec\times175\arcsec$ and took one hour to complete. The spacecraft orbital velocity and spectrograph thermal drifts were compensated by fitting the position of the \ion{Ni}{1}~279.947 nm line and subtracting its long-term trend from the signal.

\section{Results}                              \label{sec:results}

\begin{figure}
\includegraphics[scale=0.9]{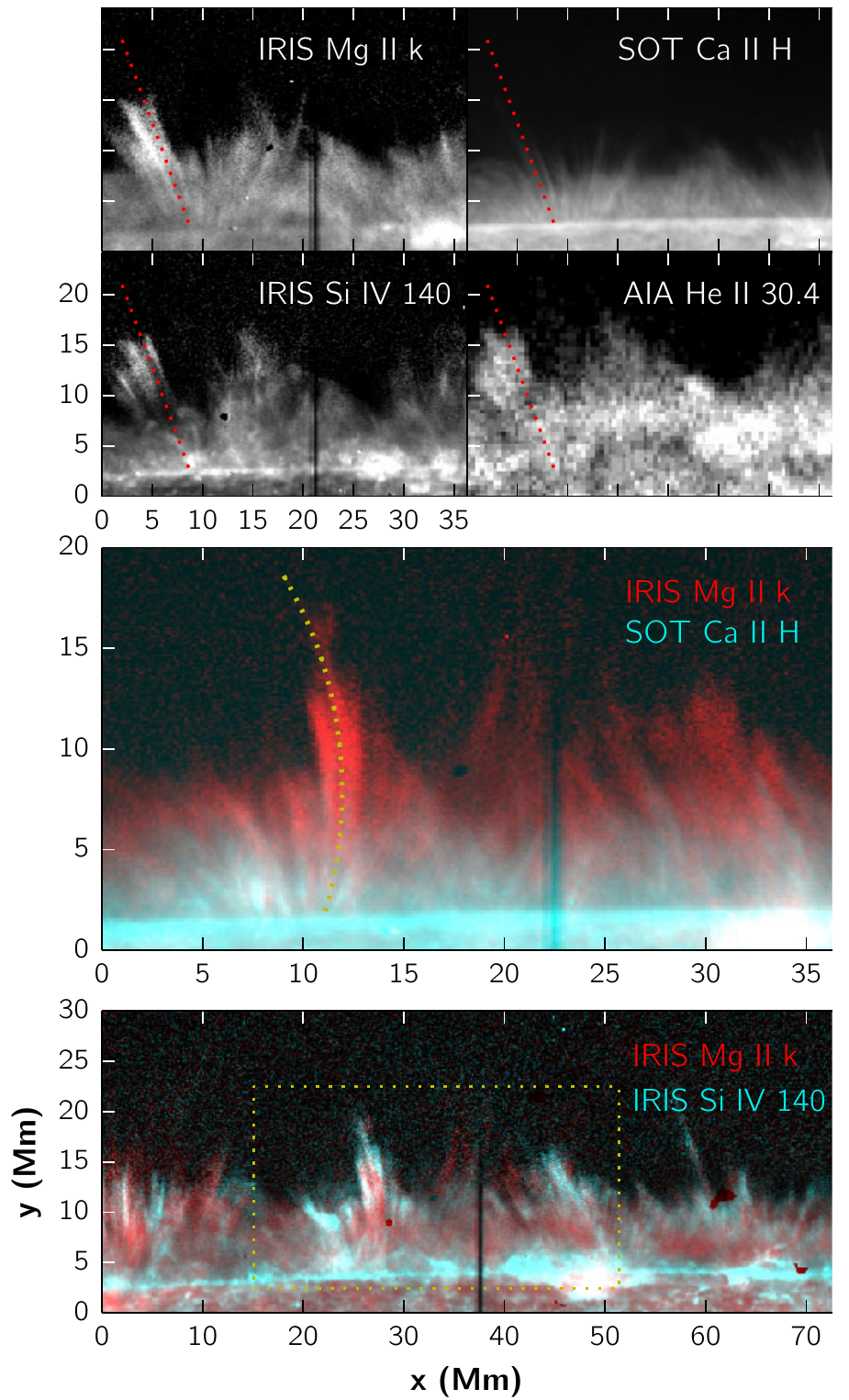} 
\caption{Quiet Sun spicule images from 2014 February 21, for two of IRIS slit-jaw filtergrams, SOT \CaH\ filtergrams, and the AIA \HeII\ 30.4~nm channel. \emph{Top four panels:} observations at 11:42:11 UT, dotted line is the region used to build the space-time diagram for spicule A in Figure \ref{fig:xt}. \emph{Middle panel:} color composite of observations at 12:19:31 UT, where the red channel corresponds to the IRIS \MgIIk\ filtergram and the green and blue channels (cyan color) to the SOT \CaH\ filtergram. The dotted line is the spicule axis (at this instant) used to build the diagram for spicule B in Figure \ref{fig:xt}. \emph{Bottom panel:} composite image using the IRIS \MgIIk\ and \SiIV\ filtergrams, for the same instant as the middle panel and a larger field. The dotted rectangle denotes the field-of-view of the middle panel. (Animated parts of this figure are available to download at \url{http://folk.uio.no/tiago/iris_spic/}.)  \label{fig:spic_panel}}
\end{figure}

\subsection{Morphology} 

In Figure~\ref{fig:spic_panel} we show a comparison between the IRIS, SOT and AIA filtergrams. The \MgIIk\ spicules are seen as a natural upward extension of the \CaH\ spicules. They are consistently taller and in several cases twice as tall. The \SiIV\ spicules show a continuation of the \MgII\ spicules and expand even farther. From the lower panel of Figure~\ref{fig:spic_panel} one can see that the \SiIV\ spicules extend just above the \MgII\ spicules virtually everywhere. Even accounting for the different spatial resolution, the IRIS spicules are broader and more nebulous than in SOT, where individual spicule ``strands'' are seen more clearly. 

Only the longest spicules are seen in the \HeII~30.4 filter, because it is opaque in higher layers that obscure the shorter spicules. \HeII\ spicules extend just above the \SiIV\ spicules and their shape is blurrier, partly because of the lower spatial resolution, and presumably in part because of the more complicated radiative transfer.

Many of the \SiIV\ spicules have a much darker bottom third or half, with the top part of the spicule noticeably brighter. This is not generally seen in \MgII\ or \CaII, and indicates that the bottom part of several spicules may not be hot enough to have significant \SiIV\ emission. This is clearly shown in the top panels of Figure~\ref{fig:spic_panel} (and in the space-time diagrams of the same spicule A in Figure~\ref{fig:xt}). Here, the \SiIV\ emission occurs primarily in the top half of the spicule, a situation that is relatively common.

\begin{figure*}
\begin{center}
\includegraphics[scale=0.74]{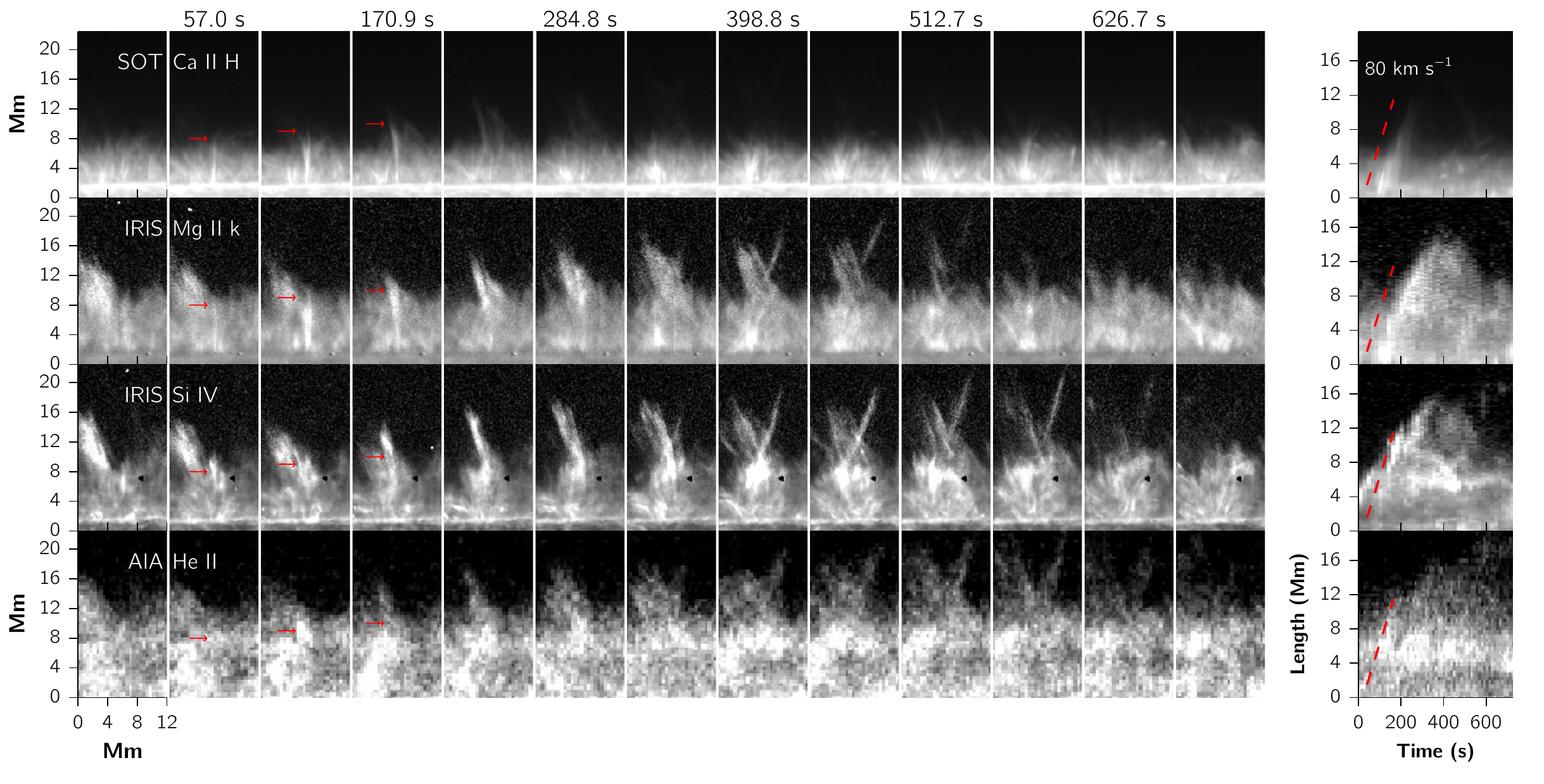} 
\end{center}
\caption{Time evolution of a spicule in different filtergrams from 2014 February 21, starting at 12:43:53 UT. The spicule of interest is indicated by the red arrows. On the right side are shown the corresponding space-time diagrams for the filtergram intensity along the axis of the spicule. The dashed red line indicates an 80~$\kms$ trajectory. %
\label{fig:spic_evo}}
\end{figure*}

\subsection{Time evolution} 
The spicules observed were mostly of type II, showing a fast rise ($> 50\:\kms$) and then fading from the \CaH\ filtergrams. In the IRIS filtergrams they have a similar upward velocity but
continue to rise after they fade in \CaII, reaching heights up to 20~Mm above the limb and often falling down later in a parabolic motion. In several cases this rise and fall is also seen in the \HeII\ 30.4~nm filtergrams. In Figure~\ref{fig:spic_evo} we show one such event where a spicule fades in \CaII\ but continues to evolve in the IRIS and AIA filters. 
This suggests that such spicules are being heated out of the \CaII\ passband, and continue evolving reaching higher layers and temperatures. This extended evolution leads to longer spicule lifetimes in the IRIS images, typically $500-800$~s. 

\begin{figure}
\begin{center}
\includegraphics[scale=0.9]{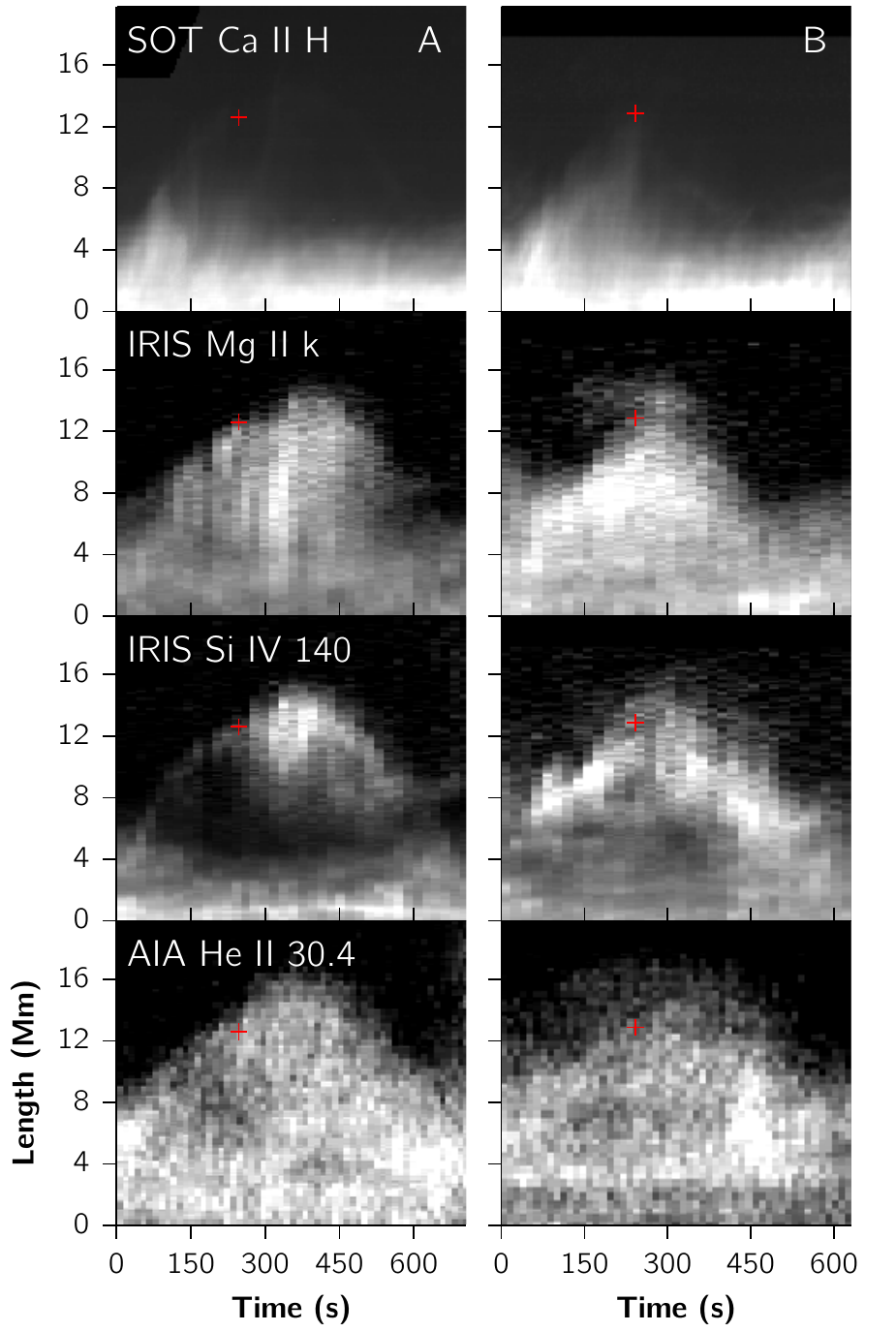} 
\end{center}
\caption{Space-time diagrams for two spicules (A and B, left and right columns respectively). Diagrams show the filtergram intensity along the axis of the spicules (following the transverse motion). The red crosses show the approximate maximum length of the spicules in the \MgIIk\ filtergram, at the instant of the images in Figure~\ref{fig:spic_panel}, where a view of these spicules is shown.\label{fig:xt}}
\end{figure}

The spicule in Figure~\ref{fig:spic_evo} is in fact a rare case of a type II spicule that is observed to fall back down in \CaII\ much later after it fades. At $t\approx512.7$~s in the intensity and in the space-time diagram it is possible to discern a faint trail from a descending spicule, roughly at the same time when the spicule is observed to fall back down by IRIS. With only the \CaII\ images it would be difficult to associate this falling material with the original spicule, but taking into account the IRIS images it becomes clear that this is most likely falling material that has cooled down and therefore is visible again in \CaII. In Figure~\ref{fig:spic_evo} there are further examples of the thermal evolution of spicules. In the first frames the later stages of a different spicule are seen on the left side of the IRIS and AIA panels, which had already faded from the \CaII\ image. At around $t\approx 350$~s a different spicule rises diagonally, towards the upper right corner. This spicule is hardly visible in \CaII\, but prominent in the other filters; it is presumably undergoing more vigorous heating.

In Figure~\ref{fig:xt} we show space-time diagrams for two other spicules, illustrating the differences in temporal evolution for different filters. 
Again, the spicules show a clear parabolic trajectory and similar lifetimes in all filters but \CaH. Spicule A shows a common evolution where \SiIV\ is much brighter at the top of the spicule, and towards the later stages of its life. A close look shows that a very faint outline of the near-parabolic trajectory of spicule A is also traced in \CaII. Spicule B also shows more \SiIV\ emission near its top. In some cases the \SiIV\ space-time diagrams look almost like half a parabola, brightening up near the later life of the spicule and coinciding with the instant where it fades in \CaII.

\begin{figure*}
\begin{center}
\includegraphics[scale=0.87]{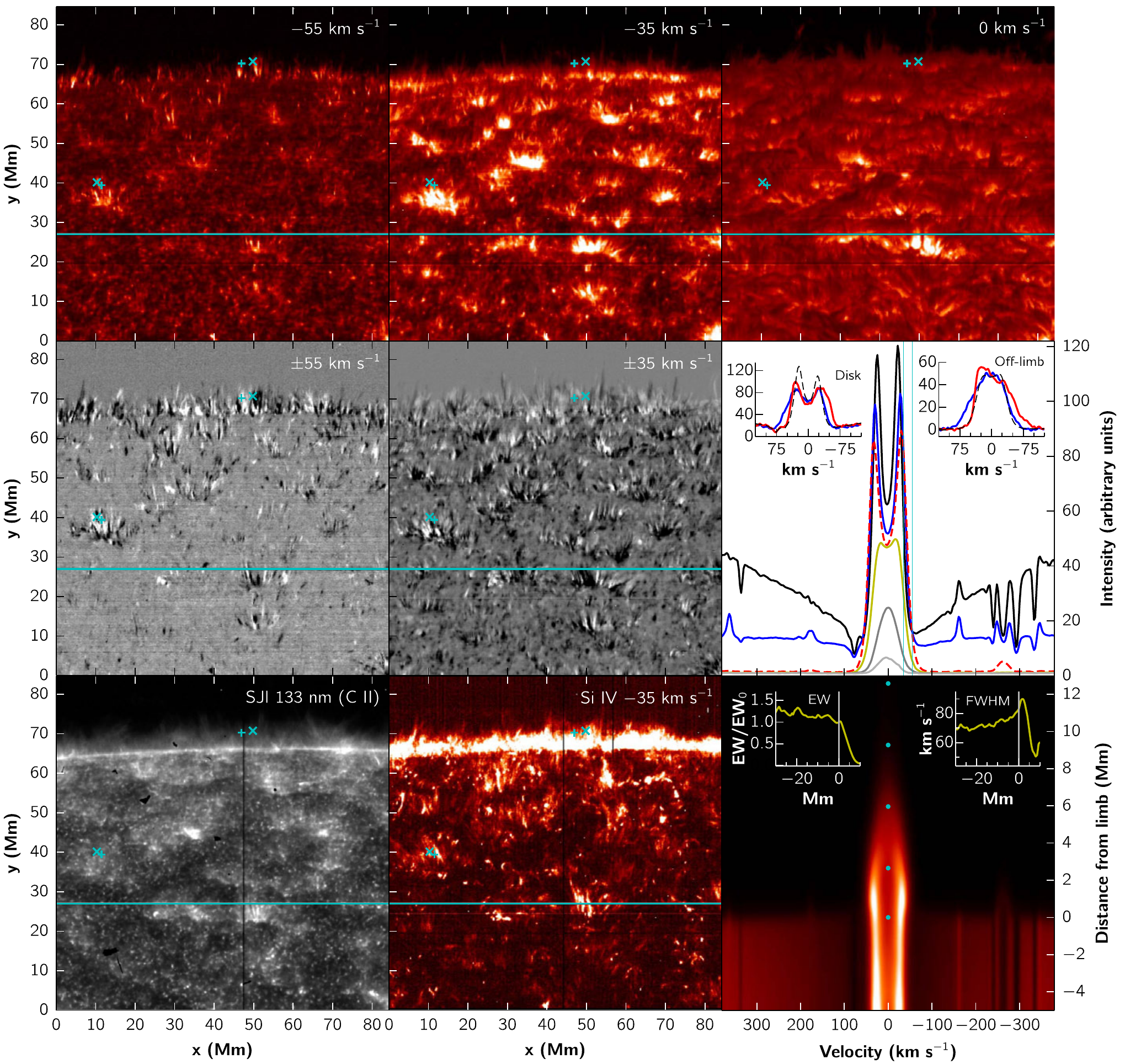} 
\end{center}
\caption{Synoptic view of a dense IRIS raster from 9 October 2013 at 13:10 UT. The size of the spatial window shown is approximately $84\times 84$~Mm ($116\arcsec \times 116\arcsec$). \emph{Top three panels:} raster intensity at fixed wavelengths or velocity shifts: $-55$, $-35$, and $0\;\kms$ from the \MgIIk\ line center. \emph{Middle left and center panels:} \MgIIk\ Dopplergrams at $\pm55$ and $\pm35\;\kms$ (black and white mean positive and negative line-of-sight velocities, respectively). \emph{Middle right:} \MgIIk\ spectra and properties, with main part showing mean spectra at several distances from the limb (\emph{black:} 16.9~Mm below the limb, \emph{blue:} at the limb, \emph{red dashed, yellow, gray and light gray:} 2.7, 6.0, 9.3, and 12.6~Mm above the limb, respectively. All but the first point are indicated by the cyan dots in the bottom right panel. The vertical cyan lines show the velocity positions at $-55$ and $-35\;\kms$. The insets show spectra at the locations of the cyan symbols in the images, compared with the mean spectrum at the same distance from the limb (black dashed). There are two sets of points: on disk [$(x, y) \approx (10, 40)$~Mm] and off the limb [$(x, y) \approx (50, 70)$~Mm]. The crosses indicate blue shifted features and are plotted with a blue line, the plus signs show red shifted features and are plotted in red. \emph{Lower left:} \CII\ 133~nm slit-jaw image taken around the middle of the raster.  \emph{Lower center:} raster intensity at $-35\;\kms$ from the \SiIV\ 139~nm line center. \emph{Lower right:} spectrogram averaged along the direction parallel to the limb. The insets show the FHWM and equivalent width (EW) as a function of the distance from the limb (EW is normalized by EW$_{0}$, the equivalent width at the limb). All images shown in a linear scale, with different saturation thresholds.\label{fig:raster}}
\end{figure*}

\subsection{Spectral rasters} 
In Figure~\ref{fig:raster} we show an overview of a dense raster. Positive velocities are defined as towards the observer. The \MgIIhk\ lines have the advantages of a wide formation range \citep{Leenaarts:Mg1, Leenaarts:Mg2, Pereira:Mg3} and a core that is much brighter than the wings. %
From Figure~\ref{fig:raster} it is clear how the disk counterparts of spicules arise naturally in bushes when observed in the wings of the k line \citep[see also][]{Beckers:1968}. As one observes in wavelengths closer to the line core there are more spicules per bush. At the line core there is more opacity from the chromospheric canopy and the spicule footpoints are no longer visible. Dopplergrams (constructed by subtracting red and blue wing images at the same wavelength offset) provide a view of the spicule line-of-sight velocities \citep[caused by the three kinds of spicule motion: upflow/downflow, transverse, and torsional, see][]{Sekse:2013aa} and how spicule numbers increase when closer to the line core.

In the middle right and bottom right panels of Figure~\ref{fig:raster} we show the mean properties of the k line as function of the distance to the limb. The equivalent width, integrated from 100 to $-100\;\kms$ from the line core, decreases by half of its limb value at about 4.5~Mm above the limb. The \MgII\ lines get broader from the disk to the limb; the FWHM increases sharply just after the limb, peaking at $\approx 1.8$~Mm above the limb and then falling steeply. In the spatially-averaged spectra the \MgII\ lines are strong even at considerable distances from the limb (here defined as zero at 280.9~nm). The k line shows a double peak until about 6~Mm above the limb, and a single-peaked emission profile is still seen beyond 12~Mm above the limb. In the insets on the middle right panel of the figure we show sample spectra from adjacent spicules, both on disk and off-limb. These spectra show the same extended wings as the Rapid Blue-shifted Events (RBEs) and Rapid Red-shifted Events (RREs) observed on disk in other lines \citep{Langangen:2008, vdVoort:2009, Sekse:2013aa}, confirming that off-limb spicules and disk RBEs/RREs have the same spectral signature.

In the bottom left panel of Figure~\ref{fig:raster} we show a \CII\ 133~nm slit-jaw image taken at a particular instant of the raster. Because of the long exposure of the raster, not all the structures seen in the slit-jaw are co-temporal with the features in the raster. Nevertheless, it is still a useful comparison. The \CII\ image shows the limb spicules and also the brighter of their disk counterparts. The fainter disk features are not resolved in the \CII\ image, but instead appear as cloudy areas \citep[see also][]{Tian:2014b}. Almost all of these cloudy features appear in the same loci of the spicule bushes from the \MgII\ raster. The \SiIV\ 140~nm slit-jaw images show essentially the same.%

In the bottom middle panel of Figure~\ref{fig:raster} one can see a TR image from the red wing of the \SiIV~139.39~nm line where spicules and their disk counterparts are still visible. Many of such spicules are a continuation of the \MgII\ spicules, in particular when seen on disk. The most obvious example is the strongest bush in the raster, at $(x, y) \approx (50, 25)$~Mm. In the \MgII\ images the spicules are just below the reference line in the figure (cyan color), and in \SiIV\ they extend more than 5~Mm above the reference line. A closer look shows that the same is true for almost all spicule bushes, even though in some cases they are noticeably fainter in \SiIV.

\section{Discussion}                           \label{sec:discussion}

IRIS brings a comprehensive new look into the evolution and origin of spicules. 
The most clear result so far is that spicules undergo thermal evolution to at least TR temperatures. The fading of type II spicules seen by SOT is not observed in the IRIS filtergrams, and is in some cases associated with increased emission in the \SiIV\ filtergrams. \MgII\ and \SiIV\ spicules continue to rise for $4-8$ Mm more, evolving for several more minutes after fading from \CaII. This strongly supports the idea that such spicules are undergoing heating and disappear from the \CaH\ passband, as previously suggested \citep{DePontieu:2009, DePontieu:2011}. Such evolution is also seen in the candidate spicule modeled by \citet{Martinez-Sykora:2013}, which disappears from \CaH\ and continues in TR filtergrams. It is somewhat puzzling that the spicules in \CaII\ are so different from \MgII, as one would expect both elements to be predominantly doubly ionized at about the same temperatures \citep[$\approx 15$~kK, see][]{Carlsson:2012}. The persistence of \MgII\ while \CaII\ disappears could indicate time-dependent ionization effects or may stem from the much higher opacity of the k line (Mg is about 18 times more abundant than Ca, and the k line is about twice as strong as the H line). This warrants further investigation.

The \SiIV\ spicules behave as one would expect with spicular heating to TR temperatures: they rise farther and are often brighter near the top.
They extend above the \MgII\ spicules in nearly all cases, and undergo essentially the same time evolution (in some rare cases spicules rise and \emph{fade} in \MgII\ while in \SiIV\ they rise and \emph{fall}). The heating to TR temperatures is confirmed by the AIA 30.4~nm channel, which shows similar dynamics and evolution to the \SiIV\ spicules, and is also seen in the spectral diagnostics. The presence of spicules in multiple filters underscores their multi-thermal nature, while the brighter tops in \SiIV\ suggest heating.

Type II spicules show parabolic space-time diagrams in the IRIS and AIA filters, but should not be confused with type I spicules. The distinction between the two types made by \citet{DePontieu:2007} was that type II were faster, much shorter lived features that faded from \CaH\ filtergrams. Our results confirm this. In addition, we now observe that this fading from \CaH\ is associated with vigorous heating to higher temperatures. The fact that type II spicules continue evolving in the IRIS filtergrams does not change the fundamental differences between the two types of spicules. Some studies such as \citet{Zhang:2012} claimed that all \CaH\ spicules behave as type I (i.e., slow and do not fade from the filtergrams), but their results could not be reproduced by \citetalias{Pereira:2012spic}, who analyzed their datasets and instead found type II spicules dominant in quiet Sun and coronal holes. The two types should not be confused.

The group behavior of spicules is clearer in the IRIS observations.
What is often perceived as one or two spicules in SOT turns out to be a much wider structure in IRIS (see Figure~\ref{fig:spic_panel}). With clearly defined multiple strands, these wide bunches of spicules have the same time evolution, and extend to areas with little or no emission in \CaII. Such group behavior plausibly contributes to the appearance of seemingly co-spatial spicules in lower resolution observations \citep{Pereira:2013spiclett}. It is likely that only the cooler strands of such groups become visible in \CaII. 
The group behavior, multi-thermal nature and apparent heating of these spicules indicate that these features are likely not driven in the same way as dynamic fibrils \citep{Hansteen:2006} even if the presence of parabolic paths is suggestive of the presence of a shock wave. Instead, the vigorous heating suggests some form of magnetic heating, perhaps as a result of magnetic reconnection \citep[as suggested by][]{DePontieu:2007}.

Detailed spectra from IRIS rasters show that the disk counterparts of quiet Sun spicules are easily seen in the wings of chromosperic and TR lines. They rise in ``bushes'' from the magnetic network as previously reported \citep{Langangen:2008, vdVoort:2009, Sekse:2012aa, Yurchyshyn:2013}, and both the on-disk and off-limb features show the same spectral signatures as the RBEs/RREs \citep[e.g.][]{Sekse:2013aa}, providing more evidence that RBEs/RREs are the disk counterparts of spicules.

Spicules as depicted by IRIS are complex and multifaceted. A pattern of multi-thermal spicules with vigorous heating is emerging, and will place constrains on much needed theoretical modeling. The field is ripe for further studies.

\acknowledgments{
   IRIS is a NASA Small Explorer mission developed and operated by LMSAL with 
   mission operations executed at NASA ARC and major  contributions to downlink
   communications funded by the NSC (Norway).
   \emph{Hinode} is a Japanese mission developed by ISAS/JAXA, with the NAOJ as domestic partner and NASA and STFC (UK) as international partners. It is operated in cooperation with ESA and NSC (Norway).
   This work was supported by the European Research Council grant No. 291058 and by NASA under contracts NNM07AA01C (\emph{Hinode}), and NNG09FA40C (IRIS). 
}

\bibliographystyle{apj}

\end{document}